\documentclass[12pt,a4paper]{article} 
\usepackage[latin1] {inputenc}
\usepackage{amsmath}
\usepackage{amsfonts} 
\usepackage{amssymb}
\usepackage{epsfig}
\numberwithin{equation}{section}
\renewcommand\P{{\rm I\kern-.25em P}}
\begin{document}
\begin{titlepage}
\hfill HD-THEP-03-63

\vspace{1.5cm}
\begin{center}
{\huge Pomeron Physics and QCD}

\vspace{2cm}
O.\ Nachtmann\footnote{e-mail: O.Nachtmann@thphys.uni-heidelberg.de}\\
\bigskip
Institut  f\"ur Theoretische Physik\\
Universit\"at Heidelberg\\
Philosophenweg 16, D-69120 Heidelberg

\vspace{2cm}
{\bf Contribution to the Ringberg Workshop on HERA physics 2003}
\end{center}

\vspace{1cm}
\begin{abstract}
We review some theoretical ideas concerning diffractive processes. We discuss the Regge
Ansatz for the pomeron and the two pomeron model. Then we present the results obtained
from nonperturbative QCD for high energy scattering. There we can extract from elastic
scattering data the parameters describing the QCD vacuum, in particular the string tension.
\end{abstract}
\end{titlepage}

\section{Introduction}
\setcounter{equation}{0}
In this contribution to the Ringberg workshop 2003 we shall discuss a topic of hadron
physics: diffractive processes at high energies. We speak of a diffractive scattering
process if one or more large rapidity gaps occur in the final state. Examples are the
reactions (see figure 1)
\begin{eqnarray}\label{1.1}
a+b&\rightarrow &X+Y,\nonumber\\
a+b&\rightarrow &X+Z+Y,
\end{eqnarray}
where $a,b$ are the incoming particles and $X,Y,Z$ can be particles
or groups of particles. Large rapidity gaps are required between $X$ and $Y$
and $X,~Y$ and $Z$. The object exchanged across the rapidity gaps is called the pomeron
(see for instance \cite{1,1a} for the history and many refences).

\begin{figure}[ht]
\begin{center}
\vspace*{.4cm}
\input{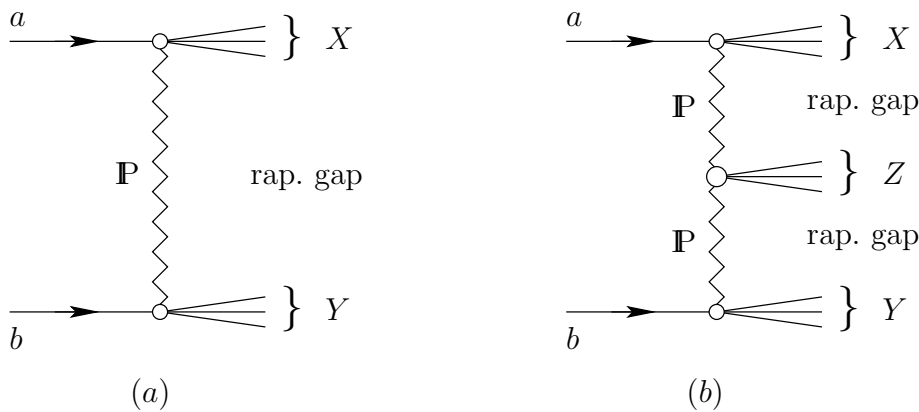}
\end{center}
\caption{Processes with (a) one and (b) two large rapidity gaps \label{figure1}}
\end{figure}

\begin{figure}[h!tb ]
\begin{center}
\input{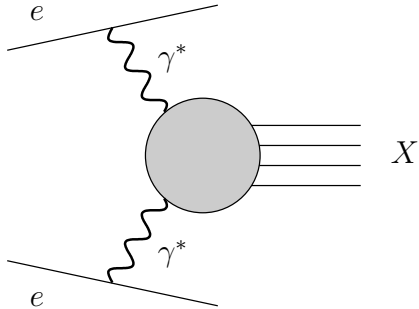}
\end{center}
\caption{
The two photon process in $e^+e^-$ collisions giving hadrons $X$ 
\label{figure2}}
\end{figure}

We can distinguish three classes of reactions involving pomerons.
\begin{itemize}
\item Soft reactions, where no large momentum transfers occur and no short distance
physics is involved. Examples are the following elastic scattering processes near the
forward direction:
\begin{eqnarray}
p+p&\rightarrow& p+p\label{1.2}\\
\pi+p&\rightarrow&\pi+p\label{1.3}\\
\gamma+p&\rightarrow&\gamma+p.\label{1.4}
\end{eqnarray}
In (\ref{1.4}) we consider real photons. In all these cases the objects scattering are
relatively large, with diameters of order $0.5$ to $1.0$ fm.
\item Semi-hard reactions, where both large and small objects and/or momentum transfers
are involved. Examples are
\begin{eqnarray}
p+p&\rightarrow & p +H+p,\label{1.5}\\
p+p&\rightarrow & p+2~ \textup{high}~ p_T ~\textup{jets}~+p\label{1.6}\\
\gamma^*+p&\rightarrow &\gamma^*+p.\label{1.7}
\end{eqnarray}
Here $H$ denotes the Higgs particle. The virtual photon $\gamma^*$ in (\ref{1.7})
is supposed to have high $Q^2$, such that it is a small object probing the large proton.
The imaginary part of the amplitude of forward virtual Compton scattering (\ref{1.7}) is,
of course, directly related to the structure functions of deep inelastic scattering.
The diffractive central production of high $p_T$ jets (\ref{1.6}) was first discussed
in \cite{2} where also the important concept of the partonic
structure of the pomeron was introduced.
\item Hard reactions, where all participating objects are small and ideally all momentum
transfers should be large. The prime example for this is $\gamma^*-\gamma^*$ scattering
with highly virtual photons
\begin{equation}\label{1.8}
\gamma^*+\gamma^*\rightarrow \gamma^*+\gamma^*.
\end{equation}
Here the observable quantity is the imaginary part of the forward scattering amplitude,
that is the total cross section for $\gamma^*\gamma^*$ going to hadrons
\begin{equation}\label{1.9}
\gamma^*+\gamma^*\rightarrow X.
\end{equation}
This can be measured for instance at $e^+e^-$ colliders, see figure \ref{figure2}.
\end{itemize}

\section{Pomerons: what are they?}
\setcounter{equation}{0}
The main question is now: what happens in a diffractive collision at high
energy, that is when a pomeron is exchanged?
Let us discuss this first for the prototype of such
collisions, elastic hadron-hadron scattering: two hadrons come
in, they interact, two hadrons go out (figure 3a). 
\begin{figure}[ht]
\begin{center}
\vspace*{.4cm}
\input{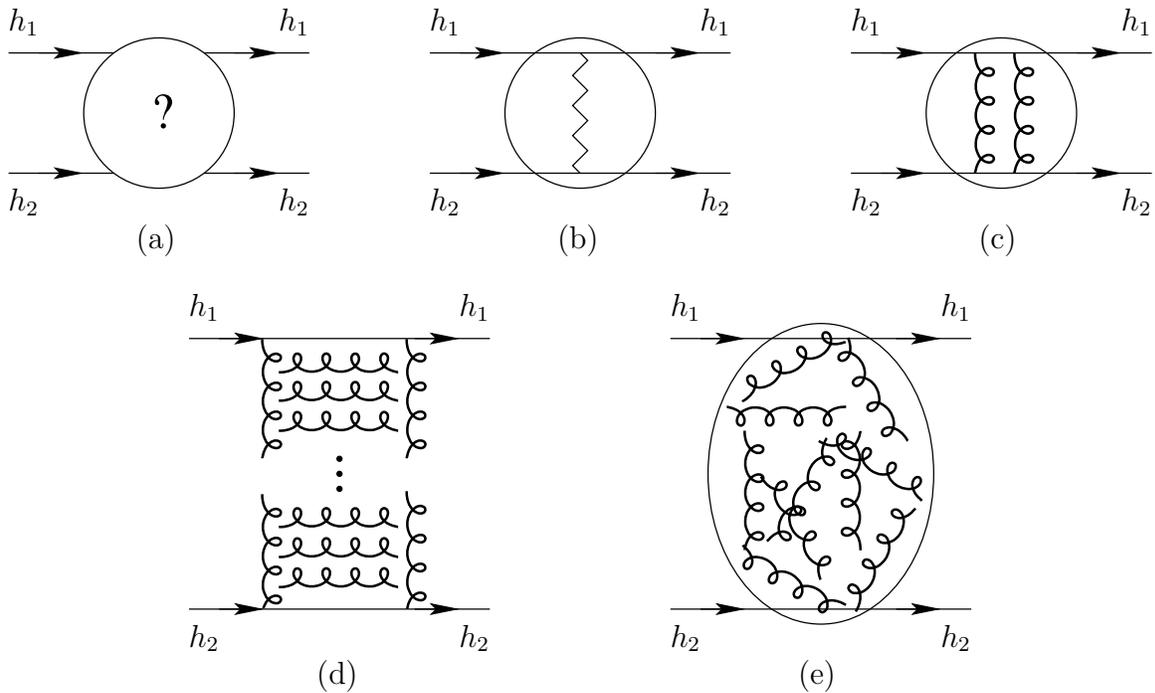}
\end{center}
\caption{Hadron-hadron scattering: (a) what happens?,
(b) phenomenological pomeron and (c) two gluon exchange, 
(d) exchange of a reggeised gluon ladder and (e) in 
the fluctuating vacuum gluon field
\label{figure3}}
\end{figure}
What happens in between? This question was, of course, already studied before the
advent of QCD. As an example of a pre-QCD approach let us discuss Regge theory which
is based on general analyticity and crossing properties of scattering amplitudes.
The Regge approach was developed in the nineteensixties and the answer provided then
to our question was: an object is exchanged,
the pomeron (figure 3b), which corresponds to the rightmost singularity in the complex
angular momentum plane. But to this date Regge theory alone
cannot predict if the pomeron is a single Regge pole, consists
of two Regge poles, is maybe a Regge cut, and so on. Nevertheless,
the assumption that the pomeron
seen in elastic hadron-hadron scattering is a simple Regge pole
works extremely well phenomenologically.

\begin{figure}[htb]
\begin{center}
\epsfxsize=.6\textwidth\epsfbox[60 580 315 770]{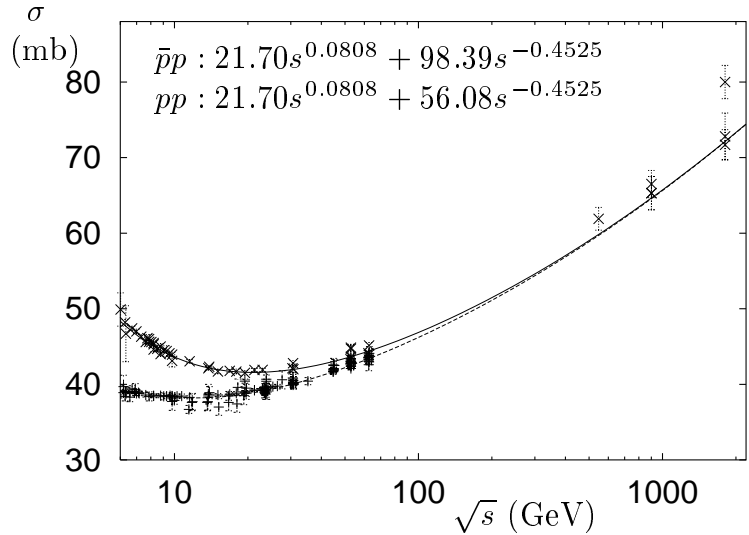}
\end{center}
\caption{Total $pp$ and $p\bar{p}$ cross sections as function of $\sqrt{s}$ (from \cite{1})}
\label{figure4}
\end{figure}

\begin{figure}[htbp]
\begin{center}
\includegraphics[width=12.5cm]{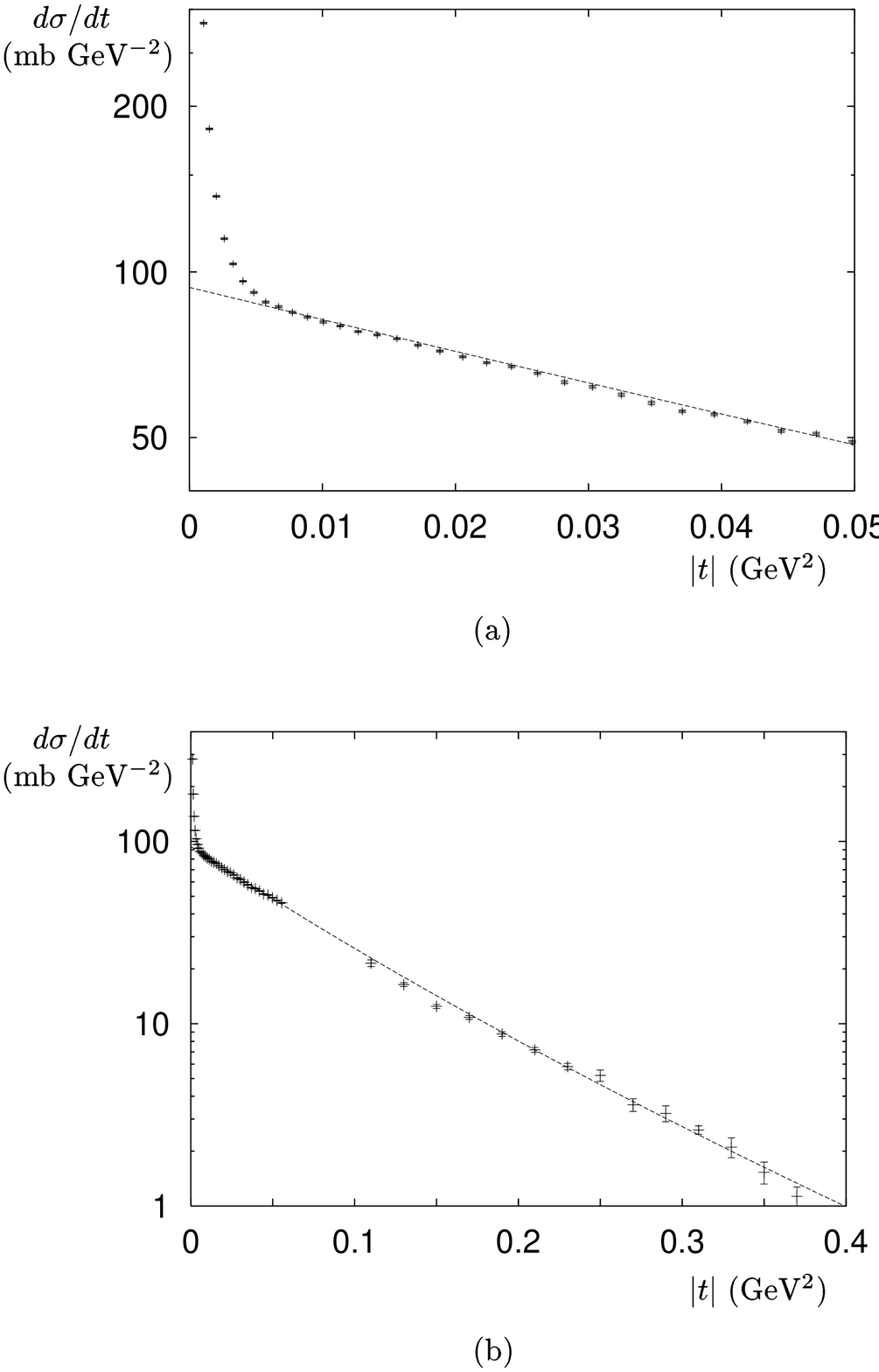}
\end{center}
\caption{$pp$ elastic scattering at $\sqrt{s} =53$ GeV (from \cite{1})}
\label{figure5}
\end{figure}

We discuss as an example the Donnachie-Landshoff (DL) Ansatz \cite{3} for the soft
pomeron in Regge theory. The DL pomeron is assumed to be effectively a simple
Regge pole. Consider as an example $p-p$ elastic scattering
\begin{eqnarray}
&&p(p_1)+p(p_2)\rightarrow p(p_3)+p(p_4),\label{2.1}\\
&&s=(p_1+p_2)^2,~t=(p_1-p_3)^2,\label{2.2}
\end{eqnarray}
where $s$ and $t$ are the c.m. energy squared and the momentum transfer squared.
The corresponding diagram is as in figure 3b, setting $h_1=h_2=p$. In the DL Ansatz
the wavy line in figure 3b can be interpreted as an effective pomeron propagator given by
\begin{equation}\label{2.3}
(-is\alpha'_{\mathbb{P}})^{\alpha_{\mathbb{P}}(t)-1}.
\end{equation}
Here $\alpha_{\mathbb{P}}(t)$ is the pomeron trajectory which is assumed to be
linear in $t$.
\begin{equation}\label{2.4}
\alpha_{\mathbb{P}}(t)=\alpha_{\mathbb{P}}(0)+\alpha'_{\mathbb{P}}t
\end{equation}
with  $\alpha_{\mathbb{P}}(0)=1+\epsilon_1$ the pomeron intercept and
$\alpha'_{\mathbb{P}}$ the slope of the pomeron trajectory. The $pp\mathbb{P}$
vertex, or Regge residue factor, is assumed to have the form
\begin{equation}\label{2.5}
-i3\beta_{\mathbb{P}}F_1(t)\gamma^\mu
\end{equation}
where $F_1(t)$ is the isoscalar Dirac electromagnetic form factor of the nucleons and
$\beta_{\mathbb{P}}$ the $\mathbb{P}$-quark coupling constant. A good representation of the
form factor is given by the dipole formula
\begin{eqnarray}\label{2.6}
F_1(t)&=&
\frac{4m^2_p-2.79t}{(4m^2_p-t)(1-t/m^2_D)^2},\nonumber\\
m^2_D&=&0.71~\textup{GeV}^2.
\end{eqnarray}
Putting everything together we find from these rules the following expression for the
scattering amplitude at large $s$
\begin{equation}\label{2.7}
\langle p(p_3,s_3),p(p_4,s_4)|{\cal{T}}|p(p_1,s_1),p(p_2,s_2)\rangle
\sim 2s\big(3\beta_{\mathbb{P}}F_1(t)\big)^2\delta_{s_3s_1}\delta_{s_4s_2}
i(-is\alpha'_{\mathbb{P}})^{\alpha_{\mathbb{P}}(t)-1}
\end{equation}
Here $s_1,\dots,s_4$ are the spin indices. This formula (\ref{2.7}) describes the data
surprisingly well and from the fits one gets the following values for the DL-pomeron
parameters:
\begin{equation}\label{2.8}
\epsilon_1=0.0808,~~\alpha'_{\mathbb{P}}=0.25~\textup{GeV}^{-2},~~
\beta_{\mathbb{P}}=1.87~\textup{GeV}^{-1}.
\end{equation}
Examples of the fits are shown in figures \ref{figure4} and
\ref{figure5} where in addition to the pomeron
also non-leading Regge exchanges are taken into account. 
Note in figure \ref{figure4} that the
Tevatron measurements of $\sigma_{tot}(p\bar{p})$ at $\sqrt{s}\simeq 1.8$ TeV give two
incompatible values. A resolution of this longstanding discrepancy would be very
welcome. The simple DL Ansatz failed, however, when applied to semi-hard diffractive
reactions. A second pomeron had to be introduced \cite{4} in order to describe the
structure functions of deep inelastic lepton-nucleon scattering, that is the total
cross section of a highly virtual photon being absorbed by a proton. This second
pomeron, the hard pomeron, was found to have an intercept much higher than the soft one:
\begin{equation}\label{2.9}
\alpha_{\mathbb{P}}(0)|_{\textup{hard}}=1+\epsilon_0,~~
\epsilon_0\cong0.44.
\end{equation}
This, of course, spoiled the simplicity of the original DL-Regge picture for the
pomeron and raised a number of questions. Is the hard pomeron also present in $pp$
scattering? Then it should show up at higher energies and maybe the LHC experiments
will be able to tell. The biggest question for a theorist is, of course, how to understand
all this in the framework of the theory of strong interactions, that is quantum
chromodynamics, QCD.

Clearly, the advent of QCD in the early
nineteenseventies changed the whole
scene in hadronic physics radically. We have since then a quantum
field theory where in principle all hadronic phenomena should be
derivable from the fundamental Lagrangian. The question arose
how to describe the pomeron in QCD. The first answers were given in
\cite{5}. These authors modeled the
pomeron as a perturbative two-gluon exchange (figure 3c).
However, this approach has problems.
Gluons in QCD are massless particles. Exchanging one of them
leads in perturbation theory
to a long-range Coulomb potential $\propto 1/r$ and a
singularity in $d\sigma/dt$ at $t=0$. Of course, in elastic
hadron-hadron scattering one must exchange at least two gluons due to
colour
conservation. But this still gives a long-range potential falling with
some power of $1/r$ and singularities in derivatives of $d\sigma/dt$
at $t=0$. From general principles we know the absence of massless hadrons to imply that
$d\sigma/dt$ is not singular at $t=0$. This is, of course,
consistent with experiment. The two-gluon exchange is the starting
point of the BFKL approach \cite{6,7} where the interaction
of the gluons is taken into account. Instead of two gluons, one
or even several gluon ladders, where the gluons themselves are
``reggeised'', are exchanged (figure 3d). Even if this is usually
called the perturbative pomeron of QCD, one should keep in mind
that this pomeron represents a very sophisticated summation of an
infinite number of diagrams in perturbation theory. For each diagram
only the leading term for high energies is kept. There is no
guaranty that the summation of the leading terms really gives the leading
term of the full theory. Also, this approach as it is based on
perturbation theory does not generate a finite length scale for the potential
and thus leads to singularities in derivatives of $d\sigma/dt$ at $t=0$.
But the BFKL pomeron should be important for
hard diffractive reactions.

For hadron-hadron scattering the total cross section and
the parameters describing $d\sigma/dt$, for instance
the slope parameter $b$,
\begin{equation}\label{2.10}
b(s)=\frac{\partial}{\partial t}\ln\frac{d\sigma}{dt}(s,t)|_{t=0},
\end{equation}
are dimensioned parameters. This fact suggested to
some authors to consider the soft pomeron as a nonperturbative object
in QCD, see \cite{8,9,10,11}, \cite{1} and references therein.
We shall treat this point of view in section 3.
In this approach hadrons scatter at small $|t|$
since the quarks in the hadrons feel the nonperturbative fluctuations
of the gluon fields in the vacuum (figure 3e). These fluctuations
are assumed to be of Gaussian nature in the stochastic vacuum
model \cite{12}. We will show that the application of this model
to high energy scattering allows us to extract from high energy data on $d\sigma/dt$
values for the vacuum parameters, the gluon condensate, the correlation
length, the non-abelian parameter and related to them the string
tension. As we shall see, the results compare well to lattice calculations and
other information on these parameters.

\section{Soft hadron reactions}
\setcounter{equation}{0}
In this section we will outline a microscopic approach towards
hadron-hadron diffractive scattering (see \cite{9,10,11}).
Consider as an example elastic scattering of two hadrons $h_1,h_2$
\begin{equation}\label{3.1}
h_1+h_2\to h_1+h_2
\end{equation}
at high energies and small momentum transfer. We will look at
reaction (\ref{3.1}) from the point of view of an
observer living in the ``femto-universe'', that is we
imagine having a microscope with resolution much better
than 1 fm for observing what happens during the collision. Of course,
we should choose an appropriate resolution
for our microscope. If we choose the resolution much too good,
we will see too many details of the internal structure of the
hadrons which are irrelevant for the reaction considered and
we will miss the essential features. The same is true if the
resolution is too poor. In \cite{9} we used a series of simple
arguments based on the uncertainty relation to estimate this
appropriate resolution.
Let $t=0$ be the nominal collision time of the hadrons in (\ref{3.1})
in the c.m. system. This is the time when the hadrons $h_1$ and $h_2$
have maximal spatial overlap. Let furthermore be $t_0/2$ the
time when, in an inelastic collision, the first produced hadrons
appear. We estimate $t_0\approx 2$ fm from the Lund model of
particle production \cite{12a}. Then the appropriate resolution, that is
the cutoff in transverse parton momenta $k_T$ of the hadronic wave
functions to be chosen for describing reaction (\ref{3.1}) in
an economical way is
\begin{equation}\label{3.2}
k^2_T\leq\sqrt s/(2t_0)
\end{equation}
where $\sqrt s$ is the c.m. energy. Modes with higher $k_T$ can be
assumed to be integrated out. With this we could argue that over the
time interval
\begin{equation}\label{3.3}
-\frac{1}{2}t_0\leq t\leq \frac{1}{2}t_0
\end{equation}
the following should hold:

\begin{itemize}
\item[(a)] The parton state of the hadrons does not change qualitatively,
that is parton annihilation and production processes can be neglected
for this time.

\item[(b)] Partons travel in essence on straight lightlike world lines.

\item[(c)] The partons undergo ``soft'' elastic scattering.
\end{itemize}

The strategy is now to study first soft parton-parton
scattering in the femto-universe. There, the relevant interaction
turns out to be mediated by the gluonic vacuum fluctuations.
It is common belief that these have a highly
nonperturbative character. In this way the nonperturbative
QCD vacuum structure enters the picture for
high energy soft hadronic reactions. Once we have solved the
problem of parton-parton scattering we have to fold the
partonic $S$-matrix with the hadronic wave functions of the
appropriate resolution (\ref{3.2}) to get the hadronic
$S$-matrix elements.

\begin{figure}[htb]
\begin{center}
\includegraphics[width=7.5cm]{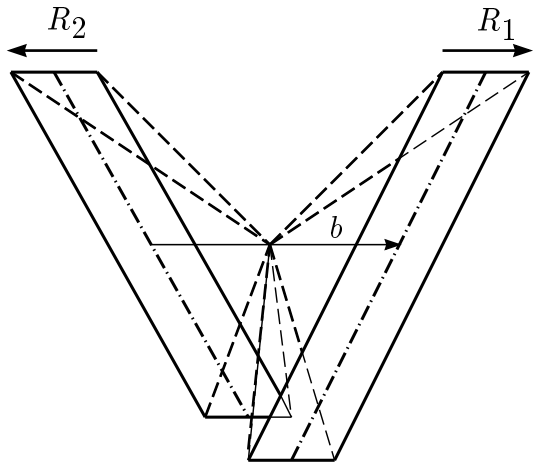}
\end{center}
\caption{Lightlike Wegner-Wilson loops in Minowski space. The sides of the loops
are given by the paths of quarks and antiquarks of the colour dipoles of sizes
$\vec{x}_T=\vec{R}_1$
and $\vec{y}_T=\vec{R}_2$. The impact parameter is $b$. (from \cite{1})}
\label{figure6}
\end{figure}

Here we can only give an outline of the various steps in this program.
Let us start by considering meson-meson scattering:
\begin{equation}\label{3.4}
M_1(P_1)+M_2(P_2)\to M_1(P_3)+M_2(P_4).
\end{equation}
We represent mesons as $q\bar{q}$ dipoles. We use standard techniques of quantum
field theory, the LSZ reduction formula and the functional integral. This allows
us in a first step to represent the amplitude for the scattering of a $q\bar{q}$
colour dipole on another $q\bar{q}$ dipole at high energies in terms of a
correlation function of two lightlike Wegner-Wilson loops (figure \ref{figure6}). 
The second
step is to fold with the meson wave functions which describe the distribution of
the colour dipoles in the mesons. The final formula reads
\begin{eqnarray}\label{3.5}
S_{fi}&=&\delta_{fi}+i(2\pi)^4\delta(P_3+P_4-P_1-P_2)T_{fi},
\nonumber\\[0.2cm]
T_{fi}&\equiv&\langle M_1(P_3),M_2(P_4)
|{\cal{T}}|M_1(P_1),M_2(P_2)\rangle\\
&=&-2is\int \textup{d}^2
b_T\textup{d}^2x_T\textup{d}^2y_Te^{i\vec q_T\cdot
\vec b_T}w_1(\vec x_T)w_2(\vec y_T)
\langle{\cal{W}}_+(\frac{1}{2}\vec b_T,\vec x_T)
{\cal{W}}_-(-\frac{1}{2}\vec b_T,\vec y_T)-1\rangle_G.
\nonumber
\end{eqnarray}
Here $s=(P_1+P_2)^2$ is the c.m. energy squared and $\vec q_T$
is the momentum transfer, which is purely transverse in the
high energy limit:
\begin{equation}\label{3.6}
\vec q_T=(\vec P_1-\vec P_3)_T
\end{equation}
and $\bar{b}_T$ is the impact parameter. Furthermore $\vec{x}_T,\vec{y_T}$ are the
transverse sizes of the colour dipoles and $w_{1,2}$ describe the distribution of the
colour dipoles in the mesons $M_{1,2}$. The symbol $\langle~\rangle_G$ denotes the
functional integration over the gluonic degrees of freedom and ${\cal{W}}$ are the
Wegner-Wilson loops.

The task is now to evaluate the functional integral
$\langle\rangle_G$ in (\ref{3.5}).
Surely we do not want to make a perturbative expansion there,
remembering our argument of sect. 2. Instead, we will turn to the
stochastic vacuum model (SVM) introduced in \cite{12} which provides a method
to calculate in a certain approximation functional integrals in the nonperturbative
domain of QCD. In the SVM the QCD vacuum is described by three parameters
\begin{itemize}
\item $G_2$, the gluon condensate introduced in \cite{13},
\item $\kappa$, the nonabelian parameter,
\item $a$, the vacuum correlation length.
\end{itemize}
The SVM relates these parameters to the string tension
\begin{equation}\label{3.7}
\sigma=\frac{32\pi\kappa G_2a^2}{81}.
\end{equation}
The SVM has been applied extensively in low energy hadron physics with considerable
success, see \cite{14} for a review. The QCD vacuum structure has also been investigated
in lattice calculations \cite{15,16}. In these one finds the values for the vacuum
parameters shown in the second column of Table 1 where the value for the string
tension is the input parameter. In the third column of the table the vacuum parameters
of the lattice results are used as input and the resulting string tension is calculated
from the SVM relation (\ref{3.7}). The outcome, $\sqrt{\sigma}=415$ MeV, is very
satisfactory and gives confidence that the SVM catches some true features of the
QCD vacuum.

\begin{table}
\begin{center}
\begin{tabular}{|l|c|c|c|}
\hline
{\rm parameter}& {\rm lattice\ calculation}& {\rm SVM}&{\rm high\ energy}\\
&{\rm quenched}& {\rm static\ pot.}&{\rm scattering}\\
\hline
&&&\\
$\sqrt\sigma$/{\rm MeV}&\underline{420}& 415&435\\
&&&\\
$G_2^{1/4}$/{\rm MeV}& 486&\underbar{486}&529\\
&&&\\
$\kappa$& 0.89&\underbar{0.89}& 0.74\\
&&&\\
$a/$fm& 0.33&\underbar{0.33}& 0.32\\[22pt]
\hline
\end{tabular}
\end{center}
\caption{
Summary of the vacuum parameters $G_2$: the gluon condensate,
$\kappa$: the nonabelian parameter, $a$: the correlation length, and of
the string tension $\sigma$, as determined with different methods. Quantities
which are underlined are input values. In the second column we list the
result of the lattice calculations \cite{16}, where $\sigma$
is taken as input. In the third column the result
of the string tension in the SVM is listed, where the vacuum parameters
$G_2, \kappa, a$ are input, see (\ref{3.7}). The fourth column lists the results for $\sigma,
\kappa, a$ obtained in \cite{18} from the data on high energy proton-proton
elastic scattering. Here the value of $G_2$ is calculated using the SVM
relation (\ref{3.7}). Errors of the numbers can be estimated to be
around 10 \%.}
\end{table}

Now we come back to high energy scattering. To evaluate the correlation function of
the two lightlike Wegner-Wilson loops in (\ref{3.5}) the SVM was continued from
Euclidean space, where it was originally formulated, to Minkowski space \cite{10}. This
opened the possibility for many applications, see for instance \cite{17}. Here I can
just cite some results as examples.

\begin{figure}[htb]
\begin{center}
\epsfxsize10cm
\epsfbox[75 550 370 770]{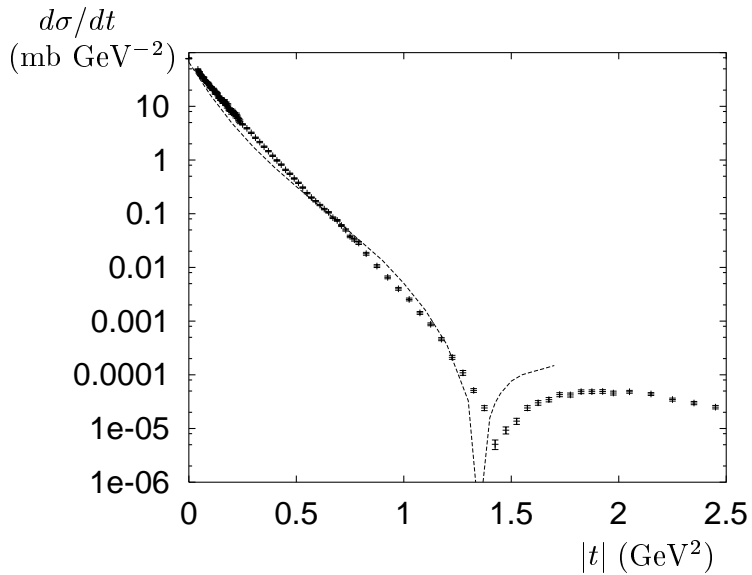}
\end{center}
\caption{Results from \cite{18} for the
elastic differential cross section for proton-proton scattering at $\sqrt{s}=23$ GeV
using the matrix cumulant expansion method (dashed line)
compared to the experimental data (from \cite{1})}
\label{figure7}
\end{figure}

The differential cross section for proton-proton elastic scattering for 
$|t|\lesssim 1$ GeV$^2$ was calculated in \cite{18}. 
The proton was considered as a quark-diquark system.
Thus the simple formula (\ref{3.5}) for meson-meson scattering could be taken over with the
diquarks replacing the antiquarks. The dipoles in the proton were assumed to have a Gaussian
distribution where the parameters can be related \cite{19} to those of
the electromagnetic form
factors. The data could then be described reasonably well with the values for the vaccum
parameters listed in column 4 of the table, see figure
\ref{figure7}. Note that the data extends
over many orders of magnitude. It was found in \cite{18} that the shape of this
$t$-distribution was quite sensitive to the vacuum parameters, in particular to $\kappa$.

We see from the table that the vacuum parameters extracted from high-energy scattering
are very well compatible with the values obtained from the lattice calculations.
In particular, the value for the square root of the string tension
$\sqrt{\sigma}=435$ MeV is well in the range obtained from the study of heavy
charmonium states. There one finds $\sqrt{\sigma}\simeq 420-440$ MeV. In our
opinion this gives support to the idea that high energy hadron-hadron scattering
is dominated by nonperturbative QCD effects.

But maybe $pp$ is special. Let us therefore look at one other result of this approach.
In figure 8 we show the pomeron contributions to various hadron-hadron total cross
sections at $\sqrt{s}=20$ GeV as function of the hadron size parameter $R_h$
divided by $R_p$.
\begin{figure}[ht]
\begin{center}
\input{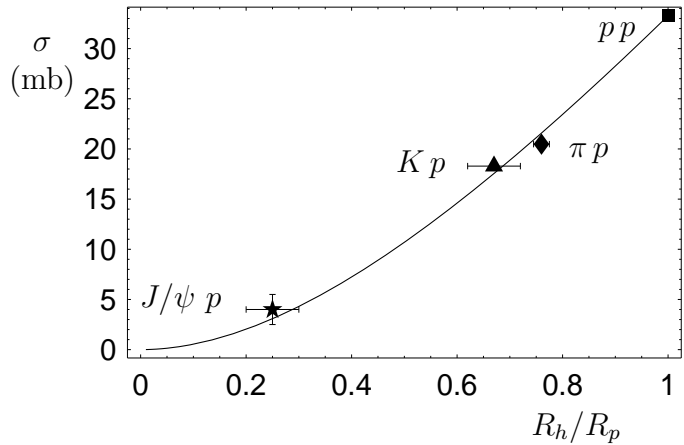}
\end{center}
\caption{The pomeron contribution at $\sqrt{s}=20$ GeV to hadron-proton total cross sections
as function of the hadron size (from \cite{1}) \label{figure8}}
\end{figure}
The curve, normalised to the $pp$ point, gives the predicted
functional dependence. Again, the results are quite satisfactory.

\section{Conclusions}
\setcounter{equation}{0}
Here we list some points which we consider important in the study of the pomeron in QCD.

\begin{itemize}
\item The phenomenological description in terms of the Regge pole Ansatz works surprisingly
well. The challenge is to understand / derive this from QCD.
\item The soft pomeron and QCD: We have developed a calculational framework based on quantum
field theoretic methods. We suggest that the soft pomeron is related to nonperturbative
features of QCD, in particular to the QCD vacuum structure. The vacuum parameters like the
string tension and the gluon condensate can be quantitatively extracted from high energy
scattering data.
\item Semi-hard diffractive phenomena, in particular the structure functions of deep
inelastic lepton-nucleon scattering are well described for instance by the two
pomeron model of Donnachie and Landshoff. Is this directly related to the usual DGLAP
evolution as suggested in \cite{20}? What is the role of the BFKL, that is the perturbative,
pomeron there? Is there saturation? Why does the dipole picture (see \cite{21,22,1}, and
the references therein) work so well? Is there a chance to make nonperturbative
calculations for the structure functions at small $x$ from first principles? Some steps
in this direction have been presented in \cite{23,24}. For hard pomeron phenomena
like the total cross section for $\gamma^*+\gamma^*\rightarrow$ hadrons presumably the
BFKL approach should work. But there are problems of large higher order corrections
\cite{25}. The experimental data \cite{26} from the LEP experiments does not give evidence for
the BFKL pomeron, so far.
\item What and where is the odderon, the $C=-1$ partner of the pomeron? See \cite{1,27}
for a discussion of this question. Experimental efforts \cite{28} to look for the
odderon should be continued and extended.
\end{itemize}

\section{Acknowledgements}
The author is grateful to the organisers of the Ringberg Workshop on
HERA Physics 2003 for inviting him to give a talk there. The atmosphere there was
stimulating and enjoyable. Many of the results discussed in this contribution could
not have been presented without the privilege the author had in being able to
collaborate with many colleagues. Special thanks in this respect are due to A. Donnachie,
H. G. Dosch and P. V. Landshoff. Finally, we thank C. Ewerz for discussions, for
reading the manuscript and for help with the figures.

\end{document}